\documentstyle[12pt,epsf]{article}

\makeatletter
\def\textfraction{0.2}
\@textmin = \textfraction\textheight
\def\fps@figure{tbp} 
\def\fps@table{tbp} 
\makeatother

\renewcommand{\textfraction}{0}

\def\gsim{\ \rlap{\raise 3pt \hbox{$>$}}{\lower 3pt \hbox{$\sim$}}\ }
\def\lsim{\ \rlap{\raise 3pt \hbox{$<$}}{\lower 3pt \hbox{$\sim$}}\ }

\begin{document}

\begin{titlepage}

\begin{flushright}
CERN-TH/98-1\\
UCHEP-98/7\\
hep-ph/9803368
\end{flushright}

\vspace{0.3cm}
\begin{center}
\boldmath
\Large\bf
Direct CP Violation in $B\to X_s\gamma$ Decays\\
as a Signature of New Physics
\unboldmath
\end{center}

\vspace{0.4cm}
\begin{center}
Alexander L. Kagan\\[0.1cm]
{\sl Department of Physics, University of Cincinnati\\
Cincinnati, Ohio 45221, USA}\\[0.3cm]
and\\[0.3cm]
Matthias Neubert\\[0.1cm]
{\sl Theory Division, CERN, CH-1211 Geneva 23, Switzerland}
\end{center}

\vspace{0.4cm}
\begin{abstract}
\vspace{0.2cm}\noindent
We argue that the observation of a sizable direct CP asymmetry $A_{\rm
CP}^{b\to s\gamma}$ in the inclusive decays $B\to X_s\gamma$ would be a
clean signal of New Physics. In the Standard Model, $A_{\rm CP}^{b\to
s\gamma}$ can be calculated reliably and is found to be below 1\% in
magnitude. In extensions of the Standard Model with new CP-violating
couplings, large CP asymmetries are possible without conflicting with
the experimental value of the branching ratio for the decays $B\to
X_s\gamma$. In particular, large asymmetries arise naturally in models
with enhanced chromo-magnetic dipole operators. Some generic examples
of such models are explored and their implications for the semileptonic
branching ratio and charm yield in $B$ decays discussed.
\end{abstract}

\vspace{0.5cm}
\centerline{(To appear in Physical Review D)}

\vfil
\noindent
CERN-TH/98-1\\
March 1998

\end{titlepage}

\section{Introduction}

Studies of rare decays of $B$ mesons have the potential to uncover the
origin of CP violation, which may lie outside the Standard Model of
strong and electroweak interactions. The measurements of several
asymmetries will make it possible to test whether the CKM mechanism of
CP violation is sufficient, or whether additional sources of CP
violation are required to describe the data. In order to achieve this
goal, it is necessary that the theoretical calculations of CP-violating
observables in terms of Standard Model parameters are, at least to a
large extent, free of hadronic uncertainties. This can be achieved, for
instance, by measuring time-dependent asymmetries in the decays of
neutral $B$ mesons into particular CP eigenstates. In many other cases,
however, the theoretical predictions for direct CP violation in
exclusive $B$ decays are obscured by large strong-interaction effects
\cite{Blok96}--\cite{At97}, which can only partly be controlled
using the approximate flavour symmetries of QCD \cite{Flei}.

Inclusive decay rates of $B$ mesons, on the other hand, can be reliably
calculated in QCD using the operator product expansion. Up to small
bound-state corrections these rates agree with the parton model
predictions for the underlying decays of the $b$ quark
\cite{Chay}--\cite{MaWe}. The possibility of observing mixing-induced
CP asymmetries in inclusive decays of neutral $B$ mesons has been
emphasized in Ref.~\cite{Ben96}. The disadvantage that the inclusive
sum over many final states partially dilutes the asymmetries is
compensated by the fact that, because of the short-distance nature of
inclusive processes, the strong phases are calculable using
quark--hadron duality. The resulting CP asymmetries are proportional to
the strong coupling constant $\alpha_s(m_b)$. The purpose of the
present paper is to study direct CP violation in the rare radiative
decays $B\to X_s\gamma$, both in the Standard Model and beyond. These
decays have already been observed experimentally, and copious data
samples will be collected at the $B$ factories. As long as the fine
structure of the photon energy spectrum is not probed locally, the
theoretical analysis relies only on the weak assumption of global
quark--hadron duality (unlike the hadronic inclusive decays considered
in Ref.~\cite{Ben96}). Also, the leading nonperturbative corrections
have been studied in detail and are well understood
\cite{Adam}--\cite{Buch}.

We perform a model-independent analysis of CP-violating effects in
$B\to X_s\gamma$ decays in terms of the effective Wilson coefficients
$C_7\equiv C_7^{\rm eff}(m_b)$ and $C_8\equiv C_8^{\rm eff}(m_b)$
multiplying the (chromo-) magnetic dipole operators
\begin{equation}
   O_7 = \frac{e\,m_b}{4\pi^2}\,\bar s_L\sigma_{\mu\nu} F^{\mu\nu}
    b_R \,, \qquad
   O_8 = \frac{g_s m_b}{4\pi^2}\,\bar s_L\sigma_{\mu\nu} G^{\mu\nu} b_R
\end{equation}
in the low-energy effective weak Hamiltonian \cite{Heff}. We will allow
for generic New Physics contributions to the coefficients $C_7$ and
$C_8$, possibly containing new CP-violating couplings. Several
extensions of the Standard Model in which new contributions to dipole
operators arise have been explored, e.g., in
Refs.~\cite{Kaga}--\cite{Bare}. We find that in the Standard Model
the direct CP asymmetry in the decays $B\to X_s\gamma$ is very small
(below 1\% in magnitude) because of a combination of CKM and GIM
suppression, both of which can be lifted in extensions of the Standard
Model. If there are new contributions to the dipole operators with
sizable weak phases, they can induce a CP asymmetry that is more than
an order of magnitude larger than in the Standard Model. We thus
propose a measurement of the inclusive CP asymmetry in the decays $B\to
X_s\gamma$ as a clean and sensitive probe of New Physics. For
simplicity, we shall not consider here the most general scenario of
having other, non-standard operators in the effective Hamiltonian.
However, we will discuss the important case of new dipole operators
involving right-handed light-quark fields, which occur, for instance,
in left--right symmetric models. The interference of these operators
with those of the standard basis, which is necessary for CP violation,
is strongly suppressed by a power of $m_s/m_b$; still, they can give
sizable contributions to CP-averaged branching ratios for rare $B$
decays.

Studies of direct CP violation in the inclusive decays $B\to X_s\gamma$
have been performed previously by several authors, both in the Standard
Model \cite{Soares} and in certain extensions of it \cite{Wolf,Asat}.
In all cases, rather small asymmetries of order a few percent or less
are obtained. Here, we generalize and extend these analyses in various
ways. Besides including some contributions to the asymmetry neglected
in previous works, we shall investigate in detail a class of New
Physics models with enhanced chromo-magnetic dipole contributions, in
which large CP asymmetries of order 10--50\% are possible and even
natural. We also perform a full next-to-leading order analysis of the
CP-averaged $B\to X_s\gamma$ branching ratio in order to derive
constraints on the parameter space of the New Physics models considered
here. For completeness, we note that CP violation has also been studied
in the related decays $B\to X_s\,\ell^+\ell^-$ \cite{Hand}, which
however have a much smaller branching ratio than the radiative decays
considered here.

\boldmath
\section{Direct CP violation in radiative $B$ decays}
\unboldmath
\label{sec:ACP}

\begin{figure}
\epsfxsize=12cm
\centerline{\epsffile{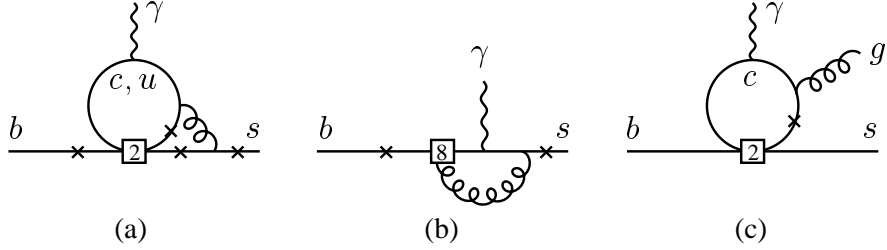}}
\centerline{\parbox{14.5cm}{\caption{\label{fig:diags}\small\sl
Diagrams for $b\to s\gamma(g)$ yielding non-trivial strong phases that
can contribute to the CP asymmetry. The crosses indicate other possible
attachments of the photon. The numbers inside the squares indicate
which operators are inserted.}}}
\end{figure}

The starting point in the calculation of the inclusive $B\to X_s\gamma$
decay rate is provided by the effective weak Hamiltonian renormalized
at the scale $\mu=m_b$ \cite{Heff}. Direct CP violation in these decays
may arise from the interference of non-trivial weak phases, contained
in CKM matrix elements or in possible New Physics contributions to the
Wilson coefficient functions, with strong phases provided by the
imaginary parts of the matrix elements of the local operators of the
effective Hamiltonian \cite{Band}. These imaginary parts first arise at
$O(\alpha_s)$ from loop diagrams containing charm quarks, light quarks
or gluons. Using the formulae of Greub et al.\ for these contributions
\cite{Greub}, we calculate at next-to-leading order the difference
$\Delta\Gamma=\Gamma(\bar B\to X_s\gamma)-\Gamma(B\to X_{\bar
s}\gamma)$ of the CP-conjugate, inclusive decay rates. The
contributions to $\Delta\Gamma$ from virtual corrections arise from
interference of the one-loop diagrams with insertions of the operators
$O_2$ and $O_8$ shown in Figure~\ref{fig:diags}(a) and (b) with the
tree-level diagram for $b\to s\gamma$ containing an insertion of the
operator $O_7$. Here $O_2=\bar s_L\gamma_\mu q_L\,\bar q_L\gamma^\mu
b_L$ with $q=c,u$ are the usual current--current operators in the
effective Hamiltonian. We find
\begin{eqnarray}
   \Delta\Gamma_{\rm virt}
   &=& \frac{G_F^2 m_b^5\alpha\alpha_s(m_b)}{18\pi^4} \nonumber\\
   &\times& \left\{ - \frac 59\,\mbox{Im}[v_u v_t^* C_2 C_7^*]
    - \left( \frac 59 - z\,v(z) \right) \mbox{Im}[v_c v_t^* C_2 C_7^*]
    - \frac{|v_t|^2}{2}\,\mbox{Im}[C_8 C_7^*] \right\} \,,
\end{eqnarray}
where $v_q=V_{qs}^* V_{qb}$ are products of CKM matrix elements,
$z=(m_c/m_b)^2$, and
\begin{equation}
   v(z) = \bigg( 5 + \ln z + \ln^2\!z - \frac{\pi^2}{3} \bigg)
   + \bigg(\! \ln^2\!z -\frac{\pi^2}{3} \bigg) z
   + \bigg( \frac{28}{9} - \frac 43 \ln z \bigg) z^2 + O(z^3) \,.
\end{equation}

\noindent
There are also contributions to $\Delta\Gamma$ from gluon
bremsstrahlung diagrams with a charm-quark loop, shown in
Figure~\ref{fig:diags}(c). They can interfere with the tree-level
diagrams for $b\to s\gamma g$ containing an insertion of $O_7$ or
$O_8$. Contrary to the virtual corrections, for which in the parton
model the photon energy is fixed to its maximum value, the gluon
bremsstrahlung diagrams lead to a non-trivial photon spectrum, and so
the results depend on the experimental lower cutoff on the photon
energy. We define a quantity $\delta$ by the requirement that $E_\gamma
> (1-\delta) E_\gamma^{\rm max}$, i.e.\ $\delta$ is the fraction of the
spectrum above the cut.\footnote{In the parton model $E_\gamma^{\rm
max}=m_b/2$ depends on the quark mass and does not agree with the
physical boundary of phase space. Later, we shall discuss how this
problem is resolved by including the effects of Fermi motion.}
We then obtain
\begin{equation}
   \Delta\Gamma_{\rm brems}
   = \frac{G_F^2 m_b^5\alpha\alpha_s(m_b)}{18\pi^4}\,z\,b(z,\delta)
   \left( \mbox{Im}[v_c v_t^* C_2 C_7^*]
   - \frac 13\,\mbox{Im}[v_c v_t^* C_2 C_8^*] \right) \,,
\end{equation}
where $b(z,\delta)=g(z,1)-g(z,1-\delta)$ with
\begin{equation}
   g(z,y) = \theta(y-4z) \left\{ (y^2-4yz+6z^2)
   \ln\!\left(\! \sqrt{\frac{y}{4z}} + \sqrt{\frac{y}{4z}-1} \,\right)
   - \frac{3y(y-2z)}{4} \sqrt{1-\frac{4z}{y}} \right\} \,.
\end{equation}
Combining the two contributions, dividing the result by the
leading-order expression for (twice) the CP-averaged inclusive decay
rate,
\begin{equation}
   \Gamma(\bar B\to X_s\gamma) + \Gamma(B\to X_{\bar s}\gamma)
   = \frac{G_F^2 m_b^5\alpha}{16\pi^4}\,|v_t C_7|^2 \,,
\label{GLO}
\end{equation}
and using the unitarity relation $v_u+v_c+v_t=0$, we find for the CP
asymmetry
\begin{eqnarray}
   A_{\rm CP}^{b\to s\gamma}(\delta)
   &=& \frac{\Gamma(\bar B\to X_s\gamma)-\Gamma(B\to X_{\bar s}\gamma)}
            {\Gamma(\bar B\to X_s\gamma)+\Gamma(B\to X_{\bar s}\gamma)}
    \Bigg|_{E_\gamma>(1-\delta) E_\gamma^{\rm max}} \nonumber\\
   &=& \frac{\alpha_s(m_b)}{|C_7|^2}\,\Bigg\{
    \frac{40}{81}\,\mbox{Im}[C_2 C_7^*]
    - \frac{8z}{9}\,\Big[ v(z) + b(z,\delta) \Big]\,
    \mbox{Im}[(1+\epsilon_s) C_2 C_7^*] \nonumber\\
   &&\hspace{1.35cm}
    \mbox{}- \frac 49\,\mbox{Im}[C_8 C_7^*]
    + \frac{8z}{27}\,b(z,\delta)\,\mbox{Im}[(1+\epsilon_s) C_2 C_8^*]
    \Bigg\} \,,
\label{ACP}
\end{eqnarray}
where
\begin{equation}
   \epsilon_s = \frac{v_u}{v_t}
   = \frac{V_{us}^* V_{ub}}{V_{ts}^* V_{tb}}
   \approx \lambda^2 (i\eta-\rho) = O(10^{-2}) \,.
\end{equation}
In the last step, we have expressed $\epsilon_s$ in terms of the
Wolfenstein parameters, with $\lambda=\sin\theta_{\rm C}\approx 0.22$
and $\rho,\eta=O(1)$. We stress that (\ref{ACP}) is an exact
next-to-leading order result. All numerical coefficients are
independent of the renormalization scheme. For consistency, the ratios
of Wilson coefficients $C_i$ must be evaluated in leading-logarithmic
order. Whereas the bremsstrahlung contributions as well as the
$C_2$--$C_8$ interference term are new, an estimate of the $C_2$--$C_7$
interference term has been obtained previously by Soares \cite{Soares},
who neglects the contribution of the function $b(z,\delta)$ and uses an
approximation for $v(z)$. The importance of the $C_8$--$C_7$
interference term for certain extensions of the Standard Model has been
stressed by Wolfenstein and Wu \cite{Wolf}, and the first correct
calculation of its coefficient can be found in Ref.~\cite{Asat}.

In the Standard Model, the Wilson coefficients take the real values
$C_2\approx 1.11$, $C_7\approx -0.31$ and $C_8\approx -0.15$. The
imaginary part of the small quantity $\epsilon_s$ is thus the only
source of CP violation. Note that all terms involving this quantity are
GIM suppressed by a power of the small ratio $z=(m_c/m_b)^2$,
reflecting the fact that there is no non-trivial weak phase difference
in the limit where $m_c=m_u=0$. Hence, the Standard Model prediction
for the CP asymmetry is suppressed by three small factors:
$\alpha_s(m_b)$ arising from the strong phases, $\sin^2\!\theta_{\rm
C}$ reflecting the CKM suppression, and $(m_c/m_b)^2$ resulting from
the GIM suppression. The numerical result for the CP asymmetry depends
on the values of the strong coupling constant and the ratio of the
heavy-quark pole masses. Throughout this work we shall take
$\alpha_s(m_b)\approx 0.214$ (corresponding to $\alpha_s(m_Z)=0.118$
and two-loop evolution down to the scale $m_b=4.8$\,GeV) and $\sqrt
z=m_c/m_b=0.29$. The sensitivity of the next-to-leading order
predictions for inclusive $B$ decay rates to theoretical uncertainties
in the values of the input parameters as well as to the choice of the
renormalization scale and scheme have been investigated by several
authors. Typically, the resulting uncertainties are of the order of
10\%. Since a discussion of such effects is not the purpose of our
study, we shall for simplicity assume fixed values of the input
parameters as quoted above. With this choice we find
\begin{equation}
   A_{\rm CP,SM}^{b\to s\gamma}(\delta)
   \approx 1.54\%\,\Big[ 1 + 0.15\,b(z,\delta) \Big]\,\eta \,,
\end{equation}
where $0\le b(z,\delta)<0.30$ depending on the value of $\delta$. With
$\eta\approx 0.2$--0.4 as suggested by phenomenological analyses
\cite{IJMP}, we find a tiny asymmetry of about 0.5\%, in agreement with
the estimate obtained in Ref.~\cite{Soares}. Expression (\ref{ACP})
applies also to the decays $B\to X_d\,\gamma$, the only difference
being that in this case the quantity $\epsilon_s$ must be replaced with
the corresponding quantity
\begin{equation}
   \epsilon_d = \frac{V_{ud}^* V_{ub}}{V_{td}^* V_{tb}}
   \approx \frac{\rho-i\eta}{1-\rho+i\eta} = O(1) \,.
\end{equation}
Therefore, in the Standard Model the CP asymmetry in $B\to X_d\,\gamma$
decays is larger by a factor $-(\lambda^2 [(1-\rho)^2+\eta^2])^{-1}
\approx -20$ than that in $B\to X_s\gamma$ decays. Note, however, that
experimentally it would be very difficult to distinguish between
inclusive $B\to X_s\gamma$ and $B\to X_d\,\gamma$ decays. If only the
sum is measured, the CP asymmetry vanishes (in the limit where
$m_s=m_d=0$), since
\begin{equation}
   \Delta\Gamma_{\rm SM}(B \to X_s\gamma)
   + \Delta\Gamma_{\rm SM}(B \to X_d\,\gamma)
   \propto \mbox{Im}\Big[ V_{ub} V_{tb}^* (V_{us}^* V_{ts}
   + V_{ud}^* V_{td}) \Big] = 0
\end{equation}
by unitarity. This has also been pointed out in Ref.~\cite{Soares}.

One might wonder whether our short-distance calculation of the CP
asymmetry in inclusive $B\to X_s\gamma$ decays could be upset by large
long-distance contributions to the decay amplitude mediated by the
current--current transitions, which could spoil quark--hadron duality.
The most important process is likely to be $B\to X_s V$ followed by
virtual $V\to\gamma$ conversion, where $V=J/\psi$ for the $b\to c\bar c
s$ transition, and $V=\rho^0,\omega^0$ for the $b\to u\bar u s$
transition. Using vector-meson dominance to estimate these effects
\cite{golowich,cheng}, we find that the largest contribution to the
asymmetry is due to $J/\psi\to\gamma$ conversion and is at most of
order 1\%, i.e.\ at the level of the prediction obtained using the
short-distance expansion. Hence, we see no reason to question the
applicability of the heavy-quark expansion to predict the inclusive CP
asymmetry.

{}From (\ref{ACP}) it is apparent that two of the suppression factors
operative in the Standard Model, $z$ and $\lambda^2$, can be avoided in
models where the effective Wilson coefficients $C_7$ and $C_8$ receive
additional contributions involving non-trivial weak phases. Much larger
CP asymmetries of $O(\alpha_s)$ then become possible. In order to
investigate such models, we may to good approximation neglect the small
quantity $\epsilon_s$ and write
\begin{equation}
   A_{\rm CP}^{b\to s\gamma}(\delta) = \frac{1}{|C_7|^2}\,\Big\{
   a_{27}(\delta)\,\mbox{Im}[C_2 C_7^*] + a_{87}\,\mbox{Im}[C_8 C_7^*]
   + a_{28}(\delta)\,\mbox{Im}[C_2 C_8^*] \Big\} \,,
\label{3terms}
\end{equation}
where
\begin{eqnarray}
   a_{27}^{(\rm p)}(\delta) &=& \alpha_s(m_b) \Bigg\{ \frac{40}{81}
    - \frac{8z}{9}\,\Big[ v(z) + b(z,\delta) \Big] \Bigg\} \,,
    \nonumber\\
   a_{87}^{(\rm p)} &=& - \frac 49\,\alpha_s(m_b) \,, \qquad
   a_{28}^{(\rm p)}(\delta) = \frac{8}{27}\,\alpha_s(m_b)\,
    z\,b(z,\delta) \,.
\label{aij}
\end{eqnarray}
The superscripts indicate that these results are obtained in the parton
model. The values of the coefficients $a_{ij}^{(\rm p)}$ are shown in
the left portion of Table~\ref{tab:aij} for three choices of the cutoff
on the photon energy: $\delta=1$ corresponding to the (unrealistic)
case of a fully inclusive measurement, $\delta=0.3$ corresponding to a
restriction to the part of the spectrum above $\approx 1.8$\,GeV, and
$\delta=0.15$ corresponding to a cutoff that removes almost all of the
background from $B$ decays into charmed hadrons. In practice, a
restriction to the high-energy part of the photon spectrum is required
for experimental reasons. Whereas the third term in (\ref{3terms}) will
generally be very small, the first two terms can give rise to sizable
effects. Since $a_{27}^{(\rm p)}$ has a rather weak dependence on
$\delta$ and $a_{87}^{(\rm p)}$ has none, the result for the CP
asymmetry is not very sensitive to the choice of the photon-energy
cutoff. Assume now that there is a New Physics contribution to $C_7$ of
similar magnitude as the Standard Model contribution, so as not to
spoil the prediction for the CP-averaged decay rate in (\ref{GLO}), but
with a non-trivial weak phase. Then the first term in (\ref{3terms})
may give a contribution of up to about 5\% in magnitude. Similarly, if
there are New Physics contributions to $C_7$ and $C_8$ such that the
ratio $C_8/C_7$ has a non-trivial weak phase, the second term may give
a contribution of up to about $10\%\times|C_8/C_7|$. In models with a
strong enhancement of $|C_8|$ with respect to its Standard Model value,
there is thus the possibility of generating very large CP asymmetries
in $B\to X_s\gamma$ decays. The relevance of the second term for
two-Higgs-doublet models, and for left--right symmetric extensions of
the Standard Model, has been explored in Refs.~\cite{Wolf,Asat}.

\begin{table}
\centerline{\parbox{14cm}{\caption{\label{tab:aij}\small\sl
Values of the coefficients $a_{ij}$ (in \%), without (left) and with
(right) Fermi motion effects included}}}
\begin{center}
\begin{tabular}{|c|ccc|cccc|}
\hline
$\delta$ & $a_{27}^{(\rm p)}$ & $a_{87}^{(\rm p)}$ & $a_{28}^{(\rm p)}$
 & $a_{27}$ & $a_{87}$ & $a_{28}$ & $E_\gamma^{\rm min}~[{\rm GeV}]$ \\
 & \multicolumn{3}{c|}{(parton model)}
 & \multicolumn{3}{c}{(with Fermi motion)} & \\
\hline\hline
1.00 & 1.06 & $-9.52$ & 0.16 & 1.06 & $-9.52$ & 0.16 & 0.00 \\
0.30 & 1.17 & $-9.52$ & 0.12 & 1.23 & $-9.52$ & 0.10 & 1.85 \\
0.15 & 1.31 & $-9.52$ & 0.07 & 1.40 & $-9.52$ & 0.04 & 2.24 \\
\hline
\end{tabular}
\end{center}
\end{table}

In our discussion so far we have neglected nonperturbative power
corrections to the inclusive decay rates. Their impact on the rate
ratio defining the CP asymmetry is expected to be very small, since
most of the corrections will cancel between the numerator and the
denominator. Potentially the most important bound-state effect is the
Fermi motion of the $b$ quark inside the $B$ meson, which determines
the shape of the photon energy spectrum in the endpoint region.
Technically, Fermi motion is included in the heavy-quark expansion by
resumming an infinite set of leading-twist corrections into a
nonperturbative ``shape function'' $F(k_+)$, which governs the
light-cone momentum distribution of the heavy quark inside the meson
\cite{shape,Dike95}. The physical decay distributions are obtained from
a convolution of parton model spectra with this function. In the
process, phase-space boundaries defined by parton model kinematics are
transformed into the proper physical boundaries defined by hadron
kinematics. For the particular case of the coefficients $a_{ij}^{(\rm
p)}(\delta)$ in (\ref{aij}), where in the parton model the parameter
$\delta$ is defined such that $E_\gamma\ge\frac 12(1-\delta) m_b$, it
can be shown that the physical coefficients $a_{ij}(\delta)$ with
$E_\gamma\ge\frac 12(1-\delta) m_B$ are given by \cite{newpaper}
\begin{equation}
   a_{ij}(\delta) =
   \frac{\int\limits_{m_B(1-\delta)-m_b}^{m_B-m_b}\!\mbox{d}k_+\,
    F(k_+)\,a_{ij}^{(\rm p)}\!\left(
    1 - \frac{m_B(1-\delta)}{m_b+k_+} \right)}
   {\int\limits_{m_B(1-\delta)-m_b}^{m_B-m_b}\!\mbox{d}k_+\,F(k_+)} \,.
\label{aijFermi}
\end{equation}
This relation is such that there is no effect if either the parton
model coefficient is independent of $\delta$, or if the limit
$\delta=1$ is taken, i.e.\ the restriction on the photon energy is
removed. Several ans\"atze for the shape function have been suggested
in the literature \cite{shape,Dike95}. For our purposes, it is
sufficient to adopt the simple form
\begin{equation}
   F(k_+) = N\,(1-x)^a e^{(1+a)x} \,;\quad
   x = \frac{k_+}{\bar\Lambda} \le 1 \,,
\end{equation}
where $\bar\Lambda=m_B-m_b$. The normalization $N$ cancels in the ratio
in (\ref{aijFermi}). The parameter $a$ can be related to the
heavy-quark kinetic energy parameter $\mu_\pi^2=-\lambda_1$
\cite{FaNe}, yielding $\mu_\pi^2=3\bar\Lambda^2/(1+a)$. In the right
portion of Table~\ref{tab:kij}, we show the values of the coefficients
$a_{ij}(\delta)$ corrected for Fermi motion, using the above ansatz
with $m_b=4.8$\,GeV and $\mu_\pi^2=0.3$\,GeV$^2$. We also give the
physical values of the minimum photon energy, $E_\gamma^{\rm min}=\frac
12(1-\delta) m_B$. The largest coefficient, $a_{87}$, is not affected
by Fermi motion, and the impact on the other two coefficients is rather
mild. As a consequence, our predictions for the CP asymmetry are very
much insensitive to bound-state effects, even if a restriction on the
high-energy part of the photon spectrum is imposed.

\boldmath
\section{Next-to-leading order corrections to $B\to X_s\gamma$}
\unboldmath

In the next section we shall explore in detail the structure of New
Physics models with a potentially large inclusive CP asymmetry. A
non-trivial constraint on such models is that they must yield an
acceptable result for the total, CP-averaged $B\to X_s\gamma$ branching
ratio, which has been measured experimentally. Taking a weighed average
of the results reported by the CLEO and ALEPH Collaborations
\cite{CLEO,ALEPH} gives $\mbox{B}(B\to X_s\gamma)=(2.5\pm 0.6)\times
10^{-4}$. We stress that this value is extracted from a measurement of
the high-energy part of the photon energy spectrum assuming that the
shape of the spectrum is as predicted by the Standard Model. For
instance, the CLEO Collaboration has measured the spectrum in the
energy range between 2.2 and 2.7\,GeV and applied a correction factor
of $0.87\pm 0.06$ in order to extrapolate to the total decay rate
\cite{private} (see Ref.~\cite{newpaper} for a critical discussion of
this treatment).

\begin{table}
\centerline{\parbox{14cm}{\caption{\label{tab:kij}\small\sl
Values of the coefficients $k_{ij}$ (in \%) with Fermi motion effects
included}}}
\begin{center}
\begin{tabular}{|cc|ccccccc|}
\hline
$\delta$ & $E_\gamma^{\rm min}~[{\rm GeV}]$ & $k_{77}$ & $k_{22}$
 & $k_{88}$ & $k_{27}$ & $k_{78}$ & $k_{28}$ & $k_{77}^{(1)}$ \\
\hline\hline
0.90 & 0.26 & 75.67 & 0.23 & 8.47 & $-14.77$ & 9.45 & $-0.04$
 & 3.47 \\
0.30 & 1.85 & 68.13 & 0.11 & 0.53 & $-16.55$ & 8.85 & $-0.01$
 & 3.86 \\
0.15 & 2.24 & 52.18 & 0.03 & 0.11 & $-13.54$ & 6.66 & $+0.00$
 & 3.15 \\
\hline
\end{tabular}
\end{center}
\end{table}

The complete theoretical prediction for the $B\to X_s\gamma$ decay rate
at next-to-leading order has been presented for the first time by
Chetyrkin et al.\ \cite{Chet}. The result for the corresponding
branching ratio is usually obtained by normalizing the radiative decay
rate to the semileptonic decay rate of $B$ mesons, thus eliminating the
strong dependence on the $b$-quark mass. We define
\begin{equation}
   \frac{\Gamma(B\to X_s\gamma)\big|_{E_\gamma>(1-\delta)
         E_\gamma^{\rm max}}}{\Gamma(B\to X_c\,e\,\bar\nu)}
   = \frac{6\alpha}{\pi f(z)}\,\left| \frac{V_{ts}^* V_{tb}}{V_{cb}}
    \right|^2 K_{\rm NLO}(\delta) \,,
\label{GNLO}
\end{equation}
where $f(z)=1-8z+8z^3-z^4-12z^2\ln z$ is a phase-space factor, and the
quantity $K_{\rm NLO}(\delta)=|C_7|^2+O(\alpha_s,1/m_b^2)$ contains the
corrections to the leading-order result. Using $\alpha^{-1}=137.036$
\cite{CzMa} and $|V_{ts}^* V_{tb}/V_{cb}|\approx 0.976$ as in
Ref.~\cite{Chet}, we get
\begin{equation}
   \mbox{B}(B\to X_s\gamma)
   \Big|_{E_\gamma>(1-\delta) E_\gamma^{\rm max}}
   \approx 2.57\times 10^{-3}\,K_{\rm NLO}(\delta)\times
   \frac{\mbox{B}(B\to X_c\,e\,\bar\nu)}{10.5\%} \,.
\end{equation}
{}From now on we shall assume the value $\mbox{B}(B\to X_c\,e\,\bar\nu)
=10.5\%$ for the semileptonic branching ratio and omit the last factor.
The current experimental situation of measurements of this quantity and
their theoretical interpretation are reviewed in Refs.~\cite{Drell,me}.
The general structure of the quantity $K_{\rm NLO}$ is
\begin{equation}
   K_{\rm NLO}(\delta) = \sum_{ \stackrel{i,j=2,7,8}{i\le j} }
   k_{ij}(\delta)\,\mbox{Re}[C_i C_j^*]
   + k_{77}^{(1)}(\delta)\,\mbox{Re}[C_7^{(1)} C_7^*] \,,
\label{KNLO}
\end{equation}
where $k_{ij}(\delta)$ are known coefficient functions depending on the
energy cutoff parameter $\delta$, and $C_7^{(1)}$ is the
next-to-leading order contribution to the Wilson coefficient $C_7^{\rm
eff}(m_b)$. In the Standard Model $C_7^{(1)}\approx 0.48$ \cite{Chet}.
Explicit expressions for the functions $k_{ij}(\delta)$, at
next-to-leading order in $\alpha_s$ and including power corrections of
order $1/m_b^2$, can be found in Ref.~\cite{newpaper}, where we correct
some mistakes in the formulae used by previous authors. (The corrected
expressions will also be given in an erratum to
Ref.~\protect\cite{Chet}.) Contrary to the case of the CP asymmetry,
the impact of Fermi motion on the partially integrated $B\to X_s\gamma$
decay rate is an important one for values of $\delta$ that are
realistic for present-day experiments. In Table~\ref{tab:kij}, we show
the values of the coefficients $k_{ij}$ corrected for Fermi motion
\cite{newpaper}, using again $m_b=4.8$\,GeV and
$\mu_\pi^2=0.3$\,GeV$^2$ for the parameters of the shape function. We
quote the results for three choices of the cutoff on the photon energy:
$\delta=0.9$ corresponding to an almost fully inclusive measurement,
and $\delta=0.3$ and 0.15 corresponding to a restriction to the
high-energy part of the photon spectrum. The choice $\delta=1$ must be
avoided because of a weak, logarithmic soft-photon divergence in the
prediction for the total $B\to X_s\gamma$ branching ratio caused by the
term proportional to $k_{88}(\delta)$. Note that with a realistic
choice of the cutoff parameter $\delta$ the coefficient $k_{88}$ of the
term proportional to $|C_8|^2$ in (\ref{KNLO}) becomes very small. This
observation will become important later on. With our choice of
parameters, we obtain in the Standard Model $\mbox{B}(B\to
X_s\gamma)=(3.3\pm 0.3)\times 10^{-4}$ for $\delta=0.9$
\cite{newpaper}, in good agreement with the results obtained in
previous analyses \cite{Chet,Buras,Ciuc}.

\begin{figure}
\epsfxsize=12cm
\centerline{\epsffile{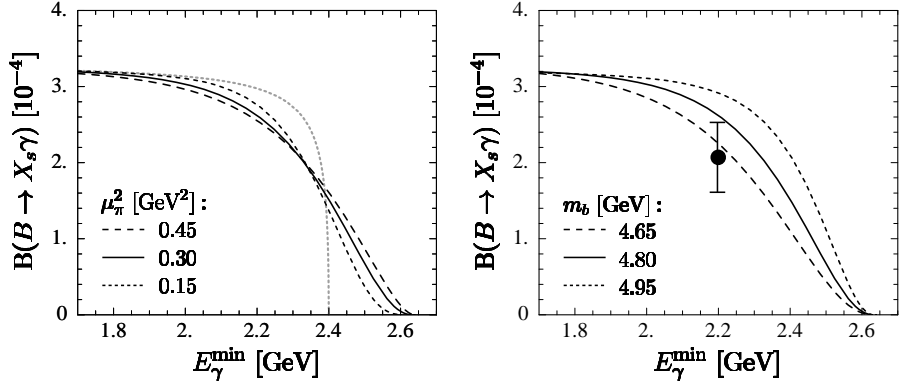}}
\centerline{\parbox{14.5cm}{\caption{\label{fig:Fermi_motion}\small\sl
Theoretical predictions for the integrated $B\to X_s\gamma$ branching
ratio for various choices of the parameters $m_b$ and $\mu_\pi^2$;
left: $\mu_\pi^2=(0.30\pm 0.15)$\,GeV$^2$ for fixed $m_b$; right:
$m_b=(4.80\pm 0.15)$\,GeV for fixed ratio $\mu_\pi^2/\bar\Lambda^2$.
The data point shows the CLEO measurement.}}}
\end{figure}

In order to illustrate the sensitivity of our results to the parameters
of the shape function, we show in Figure~\ref{fig:Fermi_motion} the
predictions for the Standard Model branching ratio as a function of the
energy cutoff $E_\gamma^{\rm min}=\frac 12(1-\delta) m_B$. In the first
plot, we keep $m_b=4.8$\,GeV fixed and compare the parton model result
(gray curve) with the results corrected for Fermi motion, using
$\mu_\pi^2=0.15$\,GeV$^2$ (short-dashed curve), 0.30\,GeV$^2$ (solid
curve), and 0.45\,GeV$^2$ (long-dashed curve). This figure illustrates
how Fermi motion fills the gap between the parton model endpoint at
$m_b/2$ and the physical endpoint\footnote{The true physical endpoint
is actually located at $[m_B^2-(m_K+m_\pi)^2]/2 m_B\approx 2.60$\,GeV,
i.e.\ slightly below $m_B/2\approx 2.64$\,GeV. Close to the endpoint,
our theoretical prediction is ``dual'' to the true spectrum in an
average sense.}
at $m_B/2$. In the second plot, we vary $m_b=4.65$\,GeV (long-dashed
curve), 4.8\,GeV (solid curve), and 4.95\,GeV (short-dashed curve),
adjusting the parameter $\mu_\pi^2$ in such a way that the ratio
$\mu_\pi^2/\bar\Lambda^2$ remains fixed. For comparison, we show the
data point $\mbox{B}(B\to X_s\gamma)=(2.04\pm 0.47)\times 10^{-4}$
obtained by the CLEO Collaboration with a cutoff at 2.2\,GeV
\cite{private}. The fact that in the CLEO analysis the cutoff is
imposed on the photon energy in the laboratory frame rather than in the
rest frame of the $B$ meson is not very important for the partially
integrated branching ratio \cite{newpaper} and will be neglected here.
Obviously, there is a rather strong dependence of the partially
integrated branching ratio on the value of the $b$-quark mass. In
particular, by choosing a low value of $m_b$ it is possible to get
agreement with the CLEO measurement without changing the prediction for
the total branching ratio. The important lesson from this investigation
is that the theoretical uncertainty in the prediction for the integral
over the high-energy part of the photon spectrum is significantly
larger than the uncertainty in the prediction of the total branching
ratio. So far, this fact has not been taken into account in the
comparison of the extrapolated experimental numbers for the total
branching ratio with theory. Ultimately, the theoretical errors may be
reduced by tuning the parameters of the shape function to fit the
measured energy spectrum; however, at present the experimental errors
are too large to make such a fit meaningful \cite{newpaper}. Below, we
shall perform our calculations for the case $\delta=0.3$ corresponding
to $E_\gamma^{\rm min}\approx 1.85$\,GeV, which is large enough to be
realistic for near-future experiments, yet low enough to be
sufficiently insensitive to the modeling of Fermi motion. As we have
pointed out before, the results for the CP asymmetry depend very little
on the choice of cutoff.

\section{CP asymmetry beyond the Standard Model}

In order to explore the implications of various New Physics scenarios
for the CP asymmetry and branching ratio in $B\to X_s\gamma$ decays it
is useful to express the Wilson coefficients $C_7=C_7^{\rm eff}(m_b)$
and $C_8=C_8^{\rm eff}(m_b)$, which are defined at the scale $m_b$, in
terms of their values at the high scale $m_W$. Using the leading-order
renormalization-group equations, one obtains
\begin{eqnarray}
   C_7 &=& \eta^\frac{16}{23}\,C_7(m_W) + \frac 83 \left(
    \eta^\frac{14}{23} - \eta^\frac{16}{23} \right) C_8(m_W)
    + \sum_{i=1}^8\,h_i\,\eta^{a_i} \,, \nonumber\\
   C_8 &=& \eta^\frac{14}{23}\,C_8(m_W)
    + \sum_{i=1}^8\,\bar h_i\,\eta^{a_i} \,,
\label{evol}
\end{eqnarray}
where $\eta=\alpha_s(m_W)/\alpha_s(m_b)\approx 0.56$, and $h_i$, $\bar
h_i$ and $a_i$ are known numerical coefficients \cite{Guidoetal,Poko}.
For the Wilson coefficients at the scale $m_W$, we write
\begin{eqnarray}
   C_7(m_W) &=& -\frac 12\,A(x_t) + C_7^{\rm new}(m_W) \,,
    \nonumber\\
   C_8(m_W) &=& -\frac 12\,D(x_t) + C_8^{\rm new}(m_W) \,,
\label{SMinitial}
\end{eqnarray}
where the first terms correspond to the leading-order Standard Model
contributions \cite{Gr90}. They are known functions of the mass ratio
$x_t=(\overline{m}_t(m_W)/m_W)^2$, which we evaluate with
$\overline{m}_t(m_W)\approx 178$\,GeV (corresponding to a pole mass of
175\,GeV). This yields $\frac 12 A(x_t)\approx 0.20$ and $\frac 12
D(x_t)\approx 0.10$. Using a similar evolution equation for the
next-to-leading coefficient $C_7^{(1)}$ \cite{Chet}, we
find\footnote{For consistency, the New Physics contributions entering
the expression for $C_7$ should be taken at next-to-leading order in
$\alpha_s(m_W)$, i.e., in the radiative decay width the corresponding
next-to-leading order New Physics matching corrections would be
accounted for through $C_7$ rather than $C_7^{(1)}$.}
\begin{eqnarray}
   C_7 &\approx& -0.31 + 0.67\,C_7^{\rm new}(m_W)
    + 0.09\,C_8^{\rm new}(m_W) \,, \nonumber\\
   C_8 &\approx& -0.15 + 0.70\,C_8^{\rm new}(m_W) \,, \nonumber\\
   C_7^{(1)} &\approx& \phantom{-}0.48 - 2.29\,C_7^{\rm new}(m_W)
    - 0.12\,C_8^{\rm new}(m_W) \,.
\label{C7C8}
\end{eqnarray}

Below, we will parametrize our results in terms of the magnitude and
phase of one of the New Physics contributions, $C_8^{\rm
new}(m_W)\equiv K_8\,e^{i\gamma_8}$ or $C_7^{\rm new}(m_W)\equiv-
K_7\,e^{i\gamma_7}$, as well as the ratio
\begin{equation}
   \xi = \frac{C_7^{\rm new}(m_W)}{Q_d\,C_8^{\rm new}(m_W)} \,,
\label{xidef}
\end{equation}
where $Q_d=-\frac 13$. A given New Physics scenario will make
predictions for these quantities at some large scale $M$. Using the
renormalization group, it is then possible to evolve these predictions
down to the scale $m_W$. At leading order, the analogues of the
relations (\ref{evol}) imply
\begin{equation}
   \xi \equiv \xi(m_W) = r\,\xi(M) - 8(1-r) \,, \qquad
   C_8^{\rm new}(m_W) = r^7\,C_8^{\rm new}(M) \,,
\end{equation}
where $r=[\alpha_s(M)/\alpha_s(m_W)]^{2/3b}$. Here $b=11-\frac 23 n_f-2
n_g$ is the first $\beta$-function coefficient, $n_f=6$ is the number
of light (with respect to the scale $M$) quark flavours, and $n_g=0,1$
denotes the number of light gluinos. For the purpose of illustration,
let us consider the three values $M=250$\,GeV, 1\,TeV and 2.5\,TeV,
which span a reasonable range of possible New Physics scales. We find
\begin{eqnarray}
   \xi &\approx& 0.98\,\xi(250\,{\rm GeV}) - 0.12 - 0.03 n_g
\nonumber\\
   &\approx& 0.97\,\xi(1\,{\rm TeV}) - 0.23 - 0.03 n_g \nonumber\\
   &\approx& 0.96\,\xi(2.5\,{\rm TeV}) - 0.29 - 0.04 n_g \,,
\label{xirela}
\end{eqnarray}
i.e.\ $\xi$ tends to be smaller than $\xi(M)$ by an amount of order
$-0.1$ to $-0.3$ depending on how close the New Physics is to the
electroweak scale. These relations will be useful for the discussion
below.

\begin{table}
\centerline{\parbox{14cm}{
\caption{\label{tab:xi}\small\sl
Ranges of $\xi(M)$ for various New Physics contributions to $C_7$ and
$C_8$, characterized by the particles in penguin diagrams}}}
\begin{center}
\begin{tabular}{|lc|lc|}
\hline
Class-1 models & $\xi(M)$ & Class-2 models & $\xi(M)$ \\
\hline\hline
neutral scalar--vectorlike quark & 1 & scalar diquark--top
 & 4.8--8.3 \\
gluino--squark ($m_{\tilde g} < 1.37 m_{\tilde q}$ )
 & $\!\!-(0.13\mbox{--}1)\!\!$ & gluino--squark ($m_{\tilde g} >
 1.37 m_{\tilde q} $) & $-(1\mbox{--}2.9)$ \\
techniscalar & $\approx-0.5$ & charged Higgs--top
 & $\!\!-(2.4\mbox{--}3.8)\!\!$ \\
 & & left--right $W$--top & $\approx -6.7$ \\
 & & Higgsino--stop & $-(2.6\mbox{--}24)$ \\
\hline
\end{tabular}
\end{center}
\end{table}

For simplicity, we shall restrict ourselves to cases where the
parameter $\xi$ in (\ref{xidef}) is real. (Otherwise there would be
even more potential for CP violation.) This happens if there is a
single dominant New Physics contribution, such as the virtual exchange
of a new heavy particle, contributing to both the magnetic and the
chromo-magnetic dipole operators. Ranges of $\xi(M)$ for several
illustrative New Physics scenarios are collected in Table~\ref{tab:xi}.
They have been obtained, for simplicity, at leading order in $\alpha_s$
and at the New Physics scale $M$ characteristic of each particular
model. With the help of the relations in (\ref{xirela}), the values of
$\xi(M)$ can be translated into the corresponding values of $\xi$,
which enter our theoretical expressions. Our aim here is not to carry
out a detailed study of each model, but to give the reader an idea of
the sizable variation that is possible in $\xi$. It is instructive to
distinguish two classes of models: those with moderate (class-1) and
those with large (class-2) values of $|\xi|$. It follows from
(\ref{C7C8}) that for small positive values of $\xi$ it is possible to
have large complex contributions to $C_8$ without affecting too much
the magnitude and phase of $C_7$, since
\begin{equation}
   \frac{C_8}{C_7}\approx\frac{0.70 K_8\,e^{i\gamma_8}-0.15}
    {(0.09-0.22\xi) K_8\,e^{i\gamma_8}-0.31} \,.
\label{C7C8rat}
\end{equation}
This is also true for small negative values of $\xi$, albeit over a
smaller region of parameter space. New Physics scenarios that have this
property belong to class-1 and have been explored in Ref.~\cite{Kaga}.
They allow for large CP asymmetries resulting from the $C_7$--$C_8$
interference term in (\ref{3terms}). Examples are penguin diagrams
containing new neutral scalars and vector-like quarks with charge
$Q_d=-\frac13$, for which $\xi(M)=1$ and hence $\xi\approx 0.8$, and
supersymmetric penguins containing light gluinos and squarks, for which
$\xi$ is negative and can be tuned by adjusting the mass ratio
$m_{\tilde g}/m_{\tilde q}$. A detailed analysis of the decays $B\to
X_s\gamma$ in the latter scenario is given in Ref.~\cite{CGG95} for the
case of real $C_7$ and $C_8$. In the table, we specifically consider
graphs with flavor off-diagonal left--right down-squark mass insertions
under the assumption that the squark masses are approximately
degenerate. The gluino and squark masses are taken to lie in the
intervals $150\,{\rm GeV}\le m_{\tilde g}\le 2.5$\,TeV and $250\,{\rm
GeV}\le m_{\tilde q}\le 2.5$\,TeV, respectively. Another example is
provided by models with techniscalars of charge $\frac 16$
\cite{Kaga,JHU,Dobrescu}, which have $\xi(M)\approx-0.5$ and hence
$\xi\approx -0.7$. In class-1 models, the magnitude of $C_8$ can be
made almost an order of magnitude larger than in the Standard Model
without spoiling the theoretical prediction for the $B\to X_s\gamma$
branching ratio.

\begin{figure}
\epsfxsize=14cm
\centerline{\epsffile{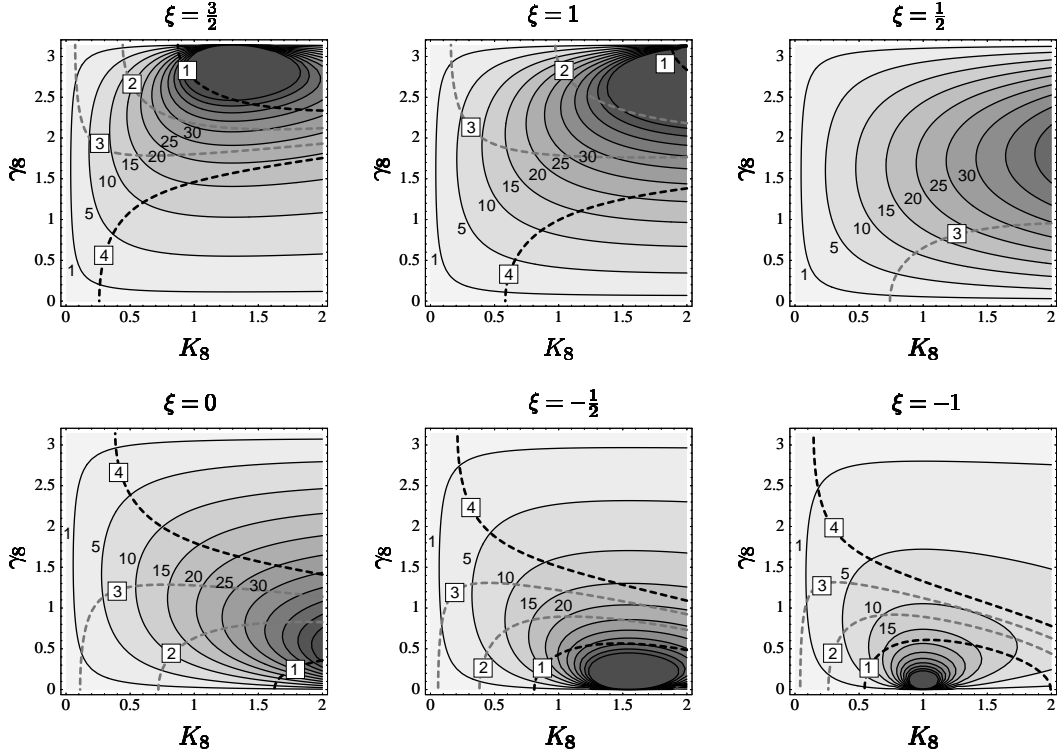}}
\centerline{\parbox{14.5cm}{\caption{\label{fig:models1}\small\sl
Contour plots for the CP asymmetry $A_{\rm CP}^{b\to s\gamma}$ for
various class-1 models}}}
\end{figure}

In Figure~\ref{fig:models1}, we show contour plots for the CP asymmetry
in the $(K_8,\gamma_8)$ plane for six different choices of $\xi$
between $\frac32$ and $-1$, assuming a cutoff $E_\gamma>1.85$\,GeV on
the photon energy (corresponding to $\delta=0.3$). We repeat that the
results for the CP asymmetry depend very little on the choice of the
cutoff. For each value of $\xi$, the plots cover the region $0\le
K_8\le 2$ and $0\le\gamma_8\le\pi$ (changing the sign of $\gamma_8$
would only change the sign of the CP asymmetry). The contour lines
refer to values of the asymmetry of 1\%, 5\%, 10\%, 15\% etc. The thick
dashed lines indicate contours where the branching ratio takes values
between $1\times 10^{-4}$ and $4\times 10^{-4}$, as indicated by the
numbers inside the squares. For comparison, we recall that the Standard
Model prediction with this choice of $\delta$ is close to $3\times
10^{-4}$, whereas the current experimental values are around $2.5\times
10^{-4}$. The main conclusion to be drawn from Figure~\ref{fig:models1}
is that in class-1 scenarios there exists great potential for sizable
CP asymmetries in a large region of parameter space. Any point to the
right of the 1\% contour for $A_{\rm CP}^{b\to s\gamma}$ cannot be
accommodated by the Standard Model. On the other hand, we see that
asymmetries of several tens of percent\footnote{We show contours only
until values $A_{\rm CP}=50\%$; for such large values, the theoretical
expression for the CP asymmetry in (\protect\ref{3terms}) would have to
be extended to higher orders to get a reliable result.} are possible in
certain extensions of the Standard Model. It is remarkable that in all
cases the regions of parameter space that yield the largest values for
the CP asymmetries are not excluded by the experimental constraint on
the CP-averaged branching ratio. This is because to have large CP
asymmetries the cross-products $C_i C_j^*$ in (\ref{3terms}) are
required to have large imaginary parts, whereas the total branching
ratio is sensitive to the real parts of these quantities. Note, in this
context, that the cutoff imposed on the photon energy strongly reduces
the size of the coefficient of the potentially dangerous term
proportional to $|C_8|^2$ in (\ref{KNLO}) and thereby helps in keeping
the prediction for the branching ratio at an acceptably low level even
for large values of $K_8$.

There are also scenarios in which the parameter $\xi$ takes on larger
negative or positive values. In such cases, it is not possible to
increase the magnitude of $C_8$ much over its Standard Model value, and
the only way to get large CP asymmetries from the $C_7$--$C_8$ or
$C_7$--$C_2$ interference terms in (\ref{3terms}) is to have $C_7$
tuned to be very small; however, this possibility is constrained by the
fact that the total $B\to X_s\gamma$ branching ratio must be of an
acceptable magnitude. That this condition starts to become a limiting
factor is already seen in the plots corresponding to $\xi=-\frac12$ and
$-1$ in Figure~\ref{fig:models1}. For even larger values of $|\xi|$,
the $C_7$--$C_8$ interference term becomes ineffective, because the
weak phase tends to cancel in the ratio $C_8/C_7$ in (\ref{C7C8rat}).
Then the $C_2$--$C_7$ interference term becomes the main source of CP
violation; however, as discussed in Section~\ref{sec:ACP}, it cannot
lead to asymmetries exceeding a level of about 5\% without violating
the constraint that the $B\to X_s\gamma$ branching ratio not be too
small. Models of this type belong to the class-2 category. Some
examples are listed in the right portion of Table~\ref{tab:xi} and can
be summarized as follows.

Models with gluino--squark loops can have large negative $\xi$ if the
ratio $m_{\tilde g}/m_{\tilde q}$ is sufficiently large. Penguin graphs
in left--right symmetric models with right-handed couplings of the $W$
boson to the top and bottom quarks and internal top-mass chirality flip
have $\xi(M)\approx\xi\approx-6.7$. Charged-Higgs--top penguins in
multi-Higgs models always have $\xi(M)<-2$ because of the charge of the
top quark. In the table graphs with internal chirality flip are
considered, with charged Higgs mass lying in the range $125\,{\rm
GeV}\le m_{H^-}\le 2.5$\,TeV (where $\xi$ increases as $ m_{H^-}$ is
increased). In general multi-Higgs models these graphs are enhanced by
a power of $m_t/m_b$ relative to their counterparts with external
chirality flip. Examples are type-3 two-Higgs-doublet models
\cite{Wolf}, left--right symmetric models \cite{Asat}, \cite{LRW1}--\cite{LRW3},
or models with additional Higgs doublets which do not acquire
significant vacuum expectation values. In all of these examples new
CP-violating phases can enter the penguin graphs, unlike in type-2
two-Higgs doublet models. Chargino--stop penguins always lead to
sizable negative values of $\xi$. For simplicity, we have considered
loops that contain a pure charged Higgsino which flips chirality. The
superpartners of new Higgs doublets with negligible vacuum expectation
values would, for example, be pure Higgsinos. The physical stop and
Higgsino masses are varied in the ranges $175\,{\rm GeV}\lsim m_{\tilde
t_1}, m_{\tilde t_2}\lsim 2.5$\,TeV and $125\,{\rm GeV}\le
m_{\tilde{h}}\le 2.5$\,TeV, respectively, under the simplifying
assumption that the stop mass matrix has equal diagonal entries, $m^2$,
and equal off-diagonal (left--right) entries, $\mu^2$, with magnitudes
satisfying $|\mu|^2\le |m m_t|$. Finally, large positive values of
$\xi$ arise from penguin graphs with a charge $-\frac13$ scalar
``diquark'' and anti-top quark in the loop. The range of values for
$\xi(M)$ quoted is again obtained for graphs with internal chirality
flip, and scalar diquark mass in the range 250\,GeV--2.5\,TeV (where
$\xi$ decreases as the scalar mass increases). In general, the phase
structure of new penguin contributions with internal and external
chirality flip will differ in the above examples; however, since the
former tend to dominate due to chiral enhancement of order $m_F/m_b$,
where $m_F$ is the mass of the heavy fermion in the loop, $\xi$ will be
real to good approximation.

\begin{figure}
\epsfxsize=14cm
\centerline{\epsffile{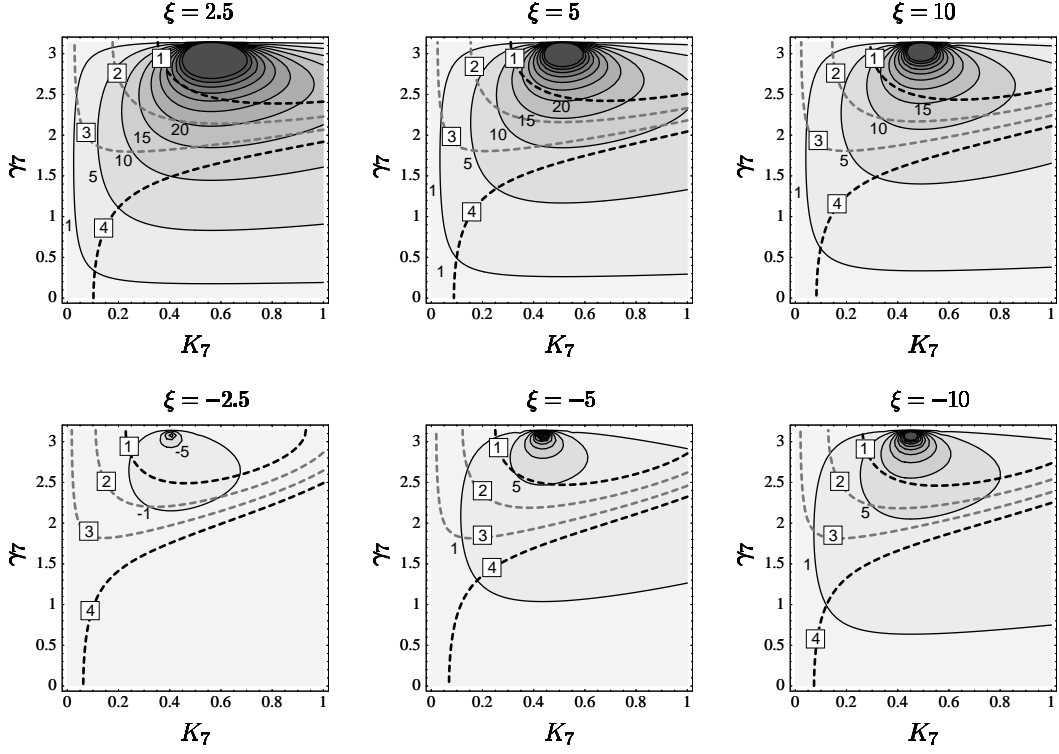}}
\centerline{\parbox{14.5cm}{\caption{\label{fig:models2}\small\sl
Contour plots for the CP asymmetry $A_{\rm CP}^{b\to s\gamma}$ for
various class-2 models}}}
\end{figure}

\begin{figure}
\epsfxsize=10cm
\centerline{\epsffile{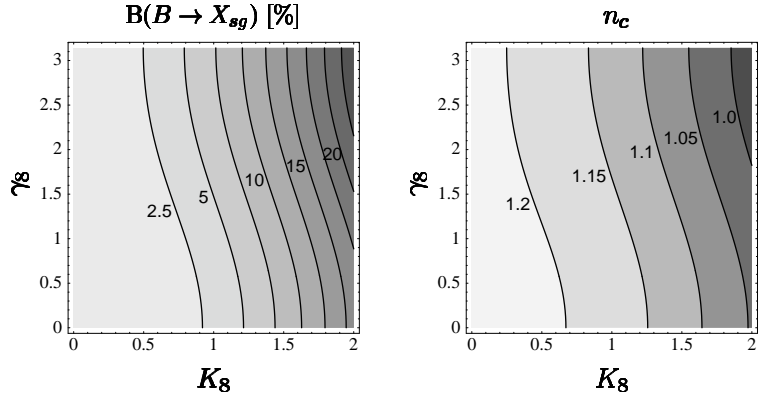}}
\centerline{\parbox{14.5cm}{\caption{\label{fig:nc}\small\sl
$B\to X_{sg}$ branching ratio (left) and charm yield in $B$ decays
(right) as a function of the parameters $K_8$ and $\gamma_8$. There is
an overall theoretical uncertainty of 6\% on the values of $n_c$.}}}
\end{figure}

For a graphical analysis of class-2 models it is convenient to choose
the magnitude and phase of the new-physics contribution $C_7^{\rm
new}(m_W)\equiv -K_7\,e^{i\gamma_7}$ as parameters, rather than $K_8$
and $\gamma_8$. The reason is that for large $|\xi|$ it becomes
increasingly unlikely that $C_8^{\rm new}(m_W)$ will be large. The
resulting plots are given in Figure \ref{fig:models2}. As before, the
dashed lines indicate the acceptable range for the $B\to X_s\gamma$
branching ratio. The branching-ratio constraint allows larger values of
$C_8$ for positive $\xi$, which explains why larger asymmetries are
attainable in this case. For example, for $\xi\approx 5$, which can be
obtained from scalar diquark--top penguins, asymmetries of 5--20\% are
seen to be consistent with the $B\to X_s\gamma$ bound. On the other
hand, for $\xi\approx-(2.5\mbox{--}5)$, which includes the
multi-Higgs-doublet models, CP asymmetries of only a few percent are
attainable, in agreement with the findings of previous authors
\cite{Wolf,Asat,newGreub}. The same is true for the left--right
symmetric $W$--top penguin, particularly if one takes into account that
$K_7\lsim 0.2$ if $m_{W_R}>1$\,TeV.

The New Physics scenarios explored in Figure~\ref{fig:models1} have the
attractive feature of a possible large enhancement of the magnitude of
the Wilson coefficient $C_8$. This has important implications for the
phenomenology of the semileptonic branching ratio and charm production
yield in $B$ decays, through enhanced production of charmless hadronic
final states induced by the $b\to s g$ flavour-changing neutral current
(FCNC) transition \cite{Kaga,CGG95,hou}. At $O(\alpha_s)$, the
theoretical expression for the $B\to X_{sg}$ decay rate is obtained
from obvious substitutions in (\ref{GLO}) to be
\begin{equation}
   \Gamma(B\to X_{sg}) = \frac{G_F^2 m_b^5\alpha_s(m_b)}{24\pi^4}\,
   |v_t C_8|^2 \,.
\label{Bsg}
\end{equation}
Normalizing this to the semileptonic rate, we obtain for the
corresponding branching ratio $\mbox{B}(B\to X_{sg})\approx 0.96\,
|C_8|^2\times {\rm B}(B\to X_c\,e\,\bar\nu)$. In the first plot in
Figure~\ref{fig:nc}, we show contours for the $B\to X_{sg}$ branching
ratio, normalized to ${\rm B}(B\to X_c\,e\,\bar\nu)=10.5\%$, in the
$(K_8,\gamma_8)$ plane. In the Standard Model, $\mbox{B}(B\to
X_{sg})\approx 0.2\%$ is very small; however, in scenarios with
$|C_8|=O(1)$ sizable values of order 10\% for this branching ratio are
possible, which simultaneously lowers the theoretical predictions for
the semileptonic branching ratio and the charm production rate $n_c$ by
a factor of $[1+\mbox{B}(B\to X_{sg})]^{-1}$. The most recent value of
$n_c$ reported by the CLEO Collaboration is $1.12\pm 0.05$
\cite{Drell}. Although the systematic errors in this measurement are
large, the result favours values of $\mbox{B}(B\to X_{sg})$ of order
10\% \cite{Raths}. This is apparent from the second plot in
Figure~\ref{fig:nc}, where we show the central theoretical prediction
for $n_c$ as a function of $K_8$ and $\gamma_8$. (There is an overall
theoretical uncertainty in the value of $n_c$ of about 6\% \cite{NeSa},
resulting from the dependence on quark masses and the renormalization
scale.) The theoretical prediction for the semileptonic branching ratio
would have the same dependence on $K_8$ and $\gamma_8$, with the
normalization ${\rm B_{SL}}=(12\pm 1)\%$ fixed at $K_8=0$ \cite{NeSa}.
A large value of $\mbox{B}(B\to X_{sg})$ could also help in
understanding the $\eta'$ yields in charmless $B$ decays
\cite{Houeta,Petrov}. For completeness, we note that the CLEO
Collaboration has recently presented a preliminary upper
limit\footnote{The limit is increased to 8.9\% if one uses the more
recent charmed baryon and charmonium yields presented in
Refs.~\protect\cite{Drell,CLEOD} and makes use of the relative
$\Lambda_c$ versus $\bar\Lambda_c$ yields given in
Ref.~\protect\cite{Cinabro}.}
on $\mbox{B}(B\to X_{sg})$ of 6.8\% (90\% CL) \cite{Thorn}. It is
therefore worth noting that large CP asymmetries of order 10--20\%
are easily attained at smaller $B\to X_{sg}$ branching ratios of a few
percent, which would nevertheless represent a marked departure from the
Standard Model prediction.

\section{Dipole operators with right-handed light quarks, and models
without CKM unitarity}

\begin{figure}
\epsfxsize=14cm
\centerline{\epsffile{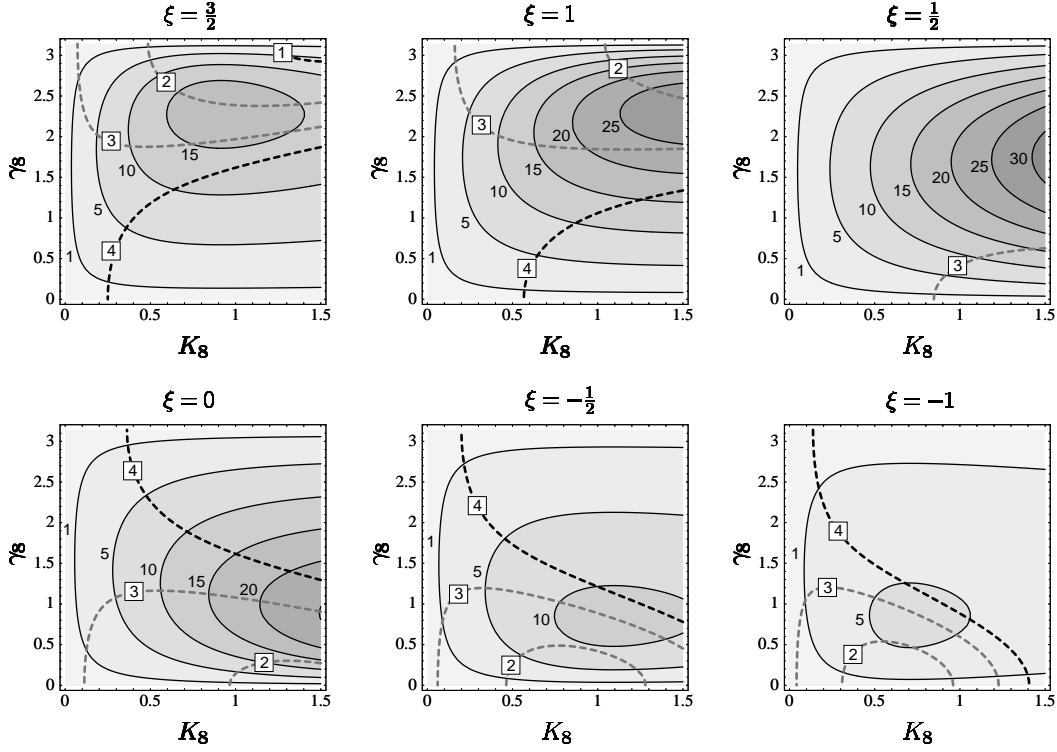}}
\centerline{\parbox{14.5cm}{\caption{\label{fig:models1LR}\small\sl
Contour plots for the CP asymmetry $A_{\rm CP}^{b\to s\gamma}$ for the
same choices of the parameter $\xi$ as in
Figure~\protect\ref{fig:models1}, but including the effect of
different-chirality operators (as explained in the text)}}}
\end{figure}

All the models listed in Table~\ref{tab:xi} can have non-standard
dipole operators involving right-handed light-quark fields. In fact, in
the absence of horizontal symmetries which impose special hierarchies
among the model parameters there is no reason why these should be any
less important than the operators of the standard basis. We therefore
briefly discuss modifications to our previous analysis in their
presence. Denoting by $C_7^R$ and $C_8^R$ the Wilson coefficients
multiplying the new operators, the expressions (\ref{3terms}),
(\ref{KNLO}) and (\ref{Bsg}) must be modified by replacing $C_i
C_j^*\to C_i C_j^* + C_i^R C_j^{R*}$ everywhere, taking however into
account that $C_2^R=0$. Note that for a single dominant New Physics
contribution the parametrization in (\ref{xidef}) for the standard
dipole operators will also be valid for the new operators, with $\xi$
taking the same real value. Then the only change in the prediction for
the CP asymmetry is that in the denominator of (\ref{3terms}) the
coefficient $|C_7|^2$ is replaced by $|C_7|^2+|C_7^R|^2$. On the other
hand, there are several new contributions to the prediction for the
total $B\to X_s\gamma$ branching ratio, as can be seen from
(\ref{KNLO}). For the purpose of illustration, let us assume that the
New Physics contributions are the same for operators of different
chirality, i.e.\ $C_i^{R,{\rm new}}(m_W)=C_i^{\rm new}(m_W)$ for
$i=7,8$. The results are shown in Figure~\ref{fig:models1LR}, where we
explore the same range of $\xi$ values as in Figure~\ref{fig:models1}.
The predictions for the $B\to X_{sg}$ branching ratio are enhanced
because $|C_8|^2$ in (\ref{Bsg}) is replaced by $|C_8|^2+|C_8^R|^2$, so
we only consider the range $0\le K_8\le 1.5$, which covers the same
values of $\mbox{B}(B\to X_{sg})$ as before. Comparing
Figures~\ref{fig:models1} and \ref{fig:models1LR}, we observe that
although there is a clear dilution of the resulting CP asymmetries
caused by the inclusion of opposite-chirality operators, there is still
plenty of parameter space in which the asymmetries are much larger than
in the Standard Model. We should also point out that, if there is more
than one significant New Physics contribution to the dipole operators,
there need not be any dilution since the product $C_8^R C_7^{R*}$ could
develop an imaginary part, thus providing an additional contribution to
the CP asymmetry.

Finally, we briefly discuss what happens in models with CKM unitarity
violation. In terms of the quantity $\Delta_s$ defined by
$v_u+v_c+(1+\Delta_s) v_t=0$, the result for the CP asymmetry in
(\ref{ACP}) generalizes to
\begin{eqnarray}
   A_{\rm CP}^{b\to s\gamma}(\delta)
   &=& \frac{\Gamma(\bar B\to X_s\gamma)-\Gamma(B\to X_{\bar s}\gamma)}
            {\Gamma(\bar B\to X_s\gamma)+\Gamma(B\to X_{\bar s}\gamma)}
    \Bigg|_{E_\gamma>(1-\delta) E_\gamma^{\rm max}} \nonumber\\
   &=& \frac{\alpha_s(m_b)}{|C_7|^2}\,\Bigg\{
    \frac{40}{81}\,\mbox{Im}[(1+\Delta_s) C_2 C_7^*]
    - \frac{8z}{9}\,\Big[ v(z) + b(z,\delta) \Big]\,
    \mbox{Im}[(1+\epsilon_s+\Delta_s) C_2 C_7^*] \nonumber\\
   &&\hspace{1.35cm}
    \mbox{}- \frac 49\,\mbox{Im}[C_8 C_7^*]
    + \frac{8z}{27}\,b(z,\delta)\,\mbox{Im}[(1+\epsilon_s
    +\Delta_s) C_2 C_8^*] \Bigg\} \,.
\label{noCKM}
\end{eqnarray}
$\Delta_s$ parametrizes the deviation from unitarity of the
3-generation CKM matrix, which could be caused, for instance, by mixing
of the known down quarks with a new isosinglet heavy quark, or by the
existence of a sequential fourth generation of quarks. In principle,
asymmetries much larger than in the Standard Model could be attained
provided that $\Delta_s$ has a significant weak phase. This reflects
the fact that the GIM suppression is no longer at work if CKM unitarity
is violated. However, we will now show that in plausible scenarios the
effect of $\Delta_s$ on the CP asymmetry is very small. In the case of
mixing with isosinglets, existing experimental limits \cite{CLEOlplm}
on the FCNC process $B\to X_s\,\ell^+\ell^-$ induced by tree-level $Z$
exchange \cite{nir} imply $\Delta_s<0.04$. The impact of non-unitarity
can therefore be safely neglected, since new contributions to the CP
asymmetry would be well below 1\%. Let us, therefore, turn to the case
of a sequential fourth generation with a new up-type quark denoted by
$t'$. As before, we will neglect the small quantity $\epsilon_s$, so
that $\mbox{Im}[\Delta_s]$ is the only source of CP violation. Then the
above expression can be rewritten in the simpler form
\begin{equation}
   A_{\rm CP}^{b\to s\gamma}(\delta)
   = a_{27}(\delta)\,\mbox{Im}\!\left[ \frac{(1+\Delta_s)C_2}{C_7}
   \right] + a_{87}\,\mbox{Im}\!\left[ \frac{C_8}{C_7} \right]
   + a_{28}(\delta)\,\mbox{Im}\!\left[ \frac{(1+\Delta_s)C_2}{C_7}
   \cdot\frac{C_8^*}{C_7^*} \right] \,.
\end{equation}
In such a scenario, the CP asymmetry is affected not only by the
non-unitarity of the 3-generation CKM matrix with
$\Delta_s=v_{t'}/v_t$ in (\ref{noCKM}), but also by the new
contributions of the $t'$ quark to the Wilson coefficients $C_7$ and
$C_8$ at the scale $m_W$. In analogy with (\ref{SMinitial}), we have
\begin{equation}
   C_7(m_W) = -\frac 12 \Big[ A(x_t) + \Delta_s A(x_{t'}) \Big]
    \,,\qquad
   C_8(m_W) = -\frac 12 \Big[ D(x_t) + \Delta_s D(x_{t'}) \Big] \,,
\end{equation}
where $x_{t'}=(\overline{m}_{t'}(m_W)/m_W)^2$. In addition, there is a
modification to the evolution equations (\ref{evol}) for the Wilson
coefficients $C_7$ and $C_8$, where now the last terms (those involving
the coefficients $h_i$ and $\bar h_i$) must be multiplied by
$-(v_c+v_u)/v_t=(1+\Delta_s)$. Taking $m_{t'}=250$\,GeV for the purpose
of illustration, we obtain $C_7\approx -0.31-0.34\Delta_s$ and
$C_8\approx -0.15-0.16\Delta_s$, i.e.\ to a good approximation we have
$C_{7,8}\approx(1+\Delta_s)C_{7,8}^{\rm SM}$. This just reflects the
fact that the functions $A(x)$ and $D(x)$ are slowly varying for $x\gg
1$. In this limit, however, all dependence on $\Delta_s$ cancels in the
expression for the CP asymmetry. As a result, there is in general not
much potential for having large CP asymmetries in models with a
sequential fourth generation. For all realistic choices of parameters,
we find asymmetries of less than 2\%, i.e.\ of a similar magnitude as
in the Standard Model.

\section{Conclusions}

We have presented a study of direct CP violation in the inclusive,
radiative decays $B\to X_s\gamma$. From a theoretical point of view,
inclusive decay rates entail the advantage of being calculable in QCD,
so that a reliable prediction for the CP asymmetry can be confronted
with data. From a practical point of view, it is encouraging that the
rare radiative decays of $B$ mesons have already been observed
experimentally, and high-statistics measurements of the corresponding
rates will be possible in the near future. We find that in the Standard
Model the CP asymmetry in $B\to X_s\gamma$ decays is strongly
suppressed by three small parameters: $\alpha_s(m_b)$ arising from the
necessity of having strong phases, $\sin^2\!\theta_{\rm C}\approx 5\%$
reflecting a CKM suppression, and $(m_c/m_b)^2\approx 8\%$ resulting
from a GIM suppression. As a result, the CP asymmetry can be safely
predicted to be of order 1\% in magnitude. This conclusion will not be
significantly modified by long-distance contributions. We have argued
that the latter two suppression factors are inoperative in extensions
of the Standard Model for which the effective Wilson coefficients $C_7$
and $C_8$ receive additional contributions involving non-trivial weak
phases. Much larger CP asymmetries of $O(\alpha_s)$ are therefore
possible in such cases.

We have presented a model-independent analysis of New Physics scenarios
in terms of the magnitudes and phases of the Wilson coefficients $C_7$
and $C_8$, finding that, indeed, sizable CP asymmetries are predicted
in large regions of parameter space. Some explicit realizations of
models with large CP asymmetries have been illustrated. In particular,
we have shown that asymmetries of 10--50\% are possible in models which
allow for a strong enhancement of the contribution from the
chromo-magnetic dipole operator. This is, in fact, quite natural unless
there is a symmetry that forbids new weak phases from entering the
coefficients $C_7$ and $C_8$. We have also shown that the predictions
for the CP asymmetry are only moderately diluted if operators involving
right-handed light-quark fields are included in the analysis. On the
other hand, we confirm the findings of previous authors regarding the
smallness of the CP asymmetry that is attainable in two-Higgs-doublet
models and in left--right symmetric models. Moreover, we find very
small effects for models in which 3-generation unitarity is violated.
Quite generally, having a large CP asymmetry is not in conflict with
the observed value for the CP-averaged $B\to X_s\gamma$ branching
ratio. On the contrary, it may even help to lower the theoretical
prediction for this quantity, and likewise for the semileptonic
branching ratio and charm multiplicity in $B$ decays, thereby bringing
these three observables closer to their experimental values.

The fact that a large inclusive CP asymmetry in $B\to X_s\gamma$ decays
is possible in many generic extensions of the Standard Model, and in a
large region of parameter space, offers the exciting possibility of
looking for a signature of New Physics in these decays using data sets
that will become available during the first period of operation of the
$B$ factories (if not existing data sets). A negative result of such a
study would impose constraints on many New Physics scenarios. A large
positive signal, on the other hand, would provide interesting clues
about the nature of physics beyond the Standard Model. In particular, a
CP asymmetry exceeding the level of 10\% would be a strong hint towards
enhanced chromo-magnetic dipole transitions caused by some new flavour
physics at a high scale.

We have restricted our analysis to the case of inclusive radiative
decays since they entail the advantage of being very clean, in the
sense that the strong-interaction phases relevant for direct CP
violation can be reliably calculated. However, if there is New Physics
that induces a large inclusive CP asymmetry in $B\to X_s\gamma$ decays,
it will inevitably also lead to sizable asymmetries in some related
processes. In particular, since we found that the inclusive CP
asymmetry remains almost unaffected if a cut on the high-energy part of
the photon energy spectrum is imposed, we expect that a large asymmetry
will persist in the exclusive mode $B\to K^*\gamma$, even though a
reliable theoretical analysis would be much more difficult because of
the necessity of calculating final-state rescattering phases
\cite{GSW95}. Still, it is worthwhile searching for a large CP
asymmetry in this channel.

Finally, it has been shown in Ref.~\cite{Atwo} that New Physics can
lead to a large time-dependent CP asymmetry in exclusive $B^0\to
K^{*0}\gamma$ decays through interference of mixing and decay. Large
direct CP violation would introduce hadronic uncertainties, thus
complicating the analysis of this effect. However, it is interesting to
note that the two phenomena are in a sense complementary in that to a
large extent they probe different New Physics contributions. We have
seen that direct CP asymmetries in radiative $B$ decays are primarily
sensitive to modifications of the Wilson coefficients of the dipole
operators with standard chirality. On the other hand, the presence of
dipole operators with right-handed light-quark fields, which are of
negligible strength in the Standard Model, is crucial for obtaining
time-dependent asymmetries, since these require both the $B^0$ and
$\bar B^0$ to be able to decay to states with the same photon helicity.

\vspace{0.3cm}
{\it Acknowledgments:\/}
We would like to thank Andrzej Buras, Lance Dixon, Gian Giudice,
Laurent Lellouch, Mikolaj Misiak, Yossi Nir and Mike Sokoloff for
helpful discussions. We are indebted to Persis Drell and Steven Glenn
for discussing aspects of the CLEO measurement of the $B\to X_s\gamma$
branching ratio. One of us (A.K.) would like to thank the CERN Theory
group for its hospitality during part of this work. A.K.~was supported
by the United States Department of Energy under Grant No.\
DE-FG02-84ER40153.

\end{document}